\documentclass[aps,epsf,twocolumn,showpacs]{revtex4}
\usepackage{amsmath}
\usepackage{epsfig}

\begin{document}

\title{Critical aspects of three-dimensional anisotropic spin-glass models}

\author{T. Papakonstantinou$^1$}

\author{N.G. Fytas$^2$}

\author{A. Malakis$^{1,2}$}

\author{I. Lelidis$^1$}

\affiliation{$^1$Department of Physics, Section of Solid State
Physics, University of Athens, Panepistimiopolis, GR 15784
Zografou, Athens, Greece}

\affiliation{$^2$ Applied Mathematics Research Centre, Coventry
University, Coventry, CV1 5FB, United Kingdom}

\date{\today}

\begin{abstract}
We study the $\pm J$ three-dimensional Ising model with a
longitudinal anisotropic bond randomness on the simple cubic
lattice. The random exchange interaction is applied only in the
$z$ direction, whereas in the other two directions, $xy$ - planes,
we consider ferromagnetic exchange. By implementing an effective
parallel tempering scheme, we outline the phase diagram of the
model and compare it to the corresponding isotropic one, as well
as to a previously studied anisotropic (transverse) case. We
present a detailed finite-size scaling analysis of the
ferromagnetic - paramagnetic and spin glass - paramagnetic
transition lines, and we also discuss the ferromagnetic - spin
glass transition regime. We conclude that the present model shares
the same universality classes with the isotropic model, but at the
symmetric point has a considerably higher transition temperature
from the spin-glass state to the paramagnetic phase. Our data for
the ferromagnetic - spin glass transition line are supporting a
forward behavior in contrast to the reentrant behavior of the
isotropic model.
\end{abstract}

 \maketitle

\section{Introduction}
\label{sec:1}

Ising spin-glass models yield phase diagrams with distinctively
complex ordered phases in three-dimensions ($d=3$). The global
phase diagram of such models includes ferromagnetic, paramagnetic,
and glassy phases, and the transitions among these are of second
order, belonging to different universality classes. An important
part of the corresponding theoretic and computational studies is
based on the Edwards-Anderson (EA) model~\cite{ea-model,BindY86}.
The EA model is defined via the Hamiltonian
\begin{equation}
\label{eq:1} \mathcal{H} = -\sum_{\langle ij
\rangle}J_{ij}s_{i}s_{j},
\end{equation}
where the summation is over nearest-neighbors, $s_{i}=\pm 1$ are
Ising spins, and $J_{ij}$ denotes the quenched uncorrelated
exchange interaction, obtained in the current work from the
following and most popular random bimodal distribution
\begin{equation}
\label{eq:2} \mathcal{P}(J_{ij}) =
p\delta(J_{ij}+1)+(1-p)\delta(J_{ij}-1).
\end{equation}

In the present paper, we consider a bimodal spin-glass model with
a spatially longitudinal anisotropic bond randomness on the simple
cubic lattice. The random exchange is applied only in the $z$
direction, whereas in the $xy$ - planes, the exchange is taken to
be ferromagnetic. Thus the disorder is longitudinal and the
interactions in the $xy$ directions are ferromagnetic. This study
follows the more general study of a spatially uniaxially
anisotropy spin-glass system considered on an hierarchical lattice
by G\"{u}ven \emph{et al}.~\cite{anis_1}, and the transverse
anisotropic model on the simple cubic lattice studied recently by
the present authors, where the random exchange was applied in the
interactions on the $xy$ - planes, while the interactions in the
$z$ direction were ferromagnetic~\cite{anis_2}. The generalized
anisotropic case studied in Ref.~\cite{anis_1} may be defined by
the Hamiltonian
\begin{equation}
\label{eq:3} \mathcal{H}^{\rm (anisotropic)}=-\sum_u\sum_{\langle
ij \rangle_u}J_{ij}^us_{i}s_{j},
\end{equation}
studied also in the current paper. Exchange interactions are
uncorrelated quenched random variables, taking values $\pm J^{xy}$
on the $xy$ - planes and $\pm J^{z}$ on the z axis. Accordingly,
the bimodal distribution of $J_{ij}^{u}$ has the general form
\begin{equation}
\label{eq:4} \mathcal{P}(J_{ij}^{u}) =
p_{u}\delta(J_{ij}^{u}+J^{u})+(1-p_{u})\delta(J_{ij}^{u}-J^{u}),
\end{equation}
where $u$ denotes the $z$ axis ($u=z$) or the $xy$ - planes ($u =
xy$), $J^{u}$ denotes the corresponding exchange interaction
strength, and $p_{u}$ are the probabilities of two neighboring
spins ($ij$) having antiferromagnetic interaction.

The above Hamiltonian includes the standard isotropic model
[Eqs.~(\ref{eq:1}) and (\ref{eq:2})], which corresponds to $J^{z}
= J^{xy} = J(=1)$ and $p_{z} = p_{xy}(=p)$. In this case, several
accurate studies have been carried out, and the critical behavior
of the corresponding ferromagnetic - paramagnetic (F - P) and spin
glass - paramagnetic (SG - P) phase transitions has been well
estimated~\cite{FG_1,hasen_fp,hasen_sg,katz_pg,jorg,campb_sg,balle_sg,
pala_sg,kawa_is,bill-11,hasen_mcp,mcp_2d,ozeki-ito,mcp_2,
doussal_harris_1,doussal_harris_2,nishimori_book,nishimori_1980,nishimori_1981,nishimori_1986}.
The ferromagnetic - spin glass (F - SG) transition line has also
been studied~\cite{FG_1}, and the multicritical point, where the
transition lines meet, located along the Nishimory line, has been
accurately
defined~\cite{hasen_mcp,mcp_2d,ozeki-ito,mcp_2,doussal_harris_1,doussal_harris_2,
nishimori_book,nishimori_1980,nishimori_1981,nishimori_1986}.

As mentioned above, our investigations concern particular cases of
the spatially anisotropic $d=3$ spin-glass system described by
Eqs.~(\ref{eq:3}) and (\ref{eq:4}). In the present study we
continue with the longitudinal anisotropic model $\{p_{z}\leq
\frac{1}{2}; p_{xy}=0\}$, whereas the previous
studies~\cite{anis_2,PT1} concerned the transverse anisotropic
model $\{p_{xy}\leq \frac{1}{2}; p_{z}=0\}$, both with $J^{z} =
J^{xy} = J(=1)$. The main motivation is the identification of
possible effects caused by the introduced anisotropy on the global
phase diagrams and the investigation of the universality aspects
of these models. This is carried out by estimating in each case
the corresponding critical exponents along the different
transition lines. Our findings are compared to both the isotropic
model and our previously studied transverse anisotropic
case~\cite{anis_2,PT1}. Similarities and differences are pointed
out and the importance of anisotropy, as well as that of the
relevant frustration features of the models, are critically
discussed.

The rest of the paper is laid out as follows: In the following
section we briefly review the Monte Carlo (MC) scheme implemented
and some details of our simulations. Section~\ref{sec:3} starts
with an outline of the finite-size scaling (FSS) schemes used in
the paper (see Section~\ref{sec:3A}), then presents our numerical
data and the estimation of the corresponding critical behavior. In
particular, in Sec.~\ref{sec:3B} the critical behavior at the F -
P transition line is estimated. For the F - P case $p_{z}=0.25$, a
detailed FSS analysis is presented that shows clearly that the
system, in this regime, belongs to the universality class of the
random Ising model, as expected. Further cases on the F - P line
are considered related to the universality class of the random
Ising model, the details of the corresponding estimations are
omitted, but the critical estimates are given in Table~\ref{tab:1}
at the end of the manuscript. In Sec.~\ref{sec:3C} the SG - P
transition is discussed with an emphasis on the symmetric case for
which a zero-temperature study is also carried out. This is
followed by Sec.~\ref{sec:3D}, where the F - SG transition is
presented. Section~\ref{sec:4} summarizes all our critical
estimates and the global phase diagrams of the isotropic EA model
and the two anisotropic variants $p_{z}=0; p_{xy}\leq \frac{1}{2}$
and $p_{z}\leq \frac{1}{2}; p_{xy}=0$ are sketched and discussed,
pointing out similarities, as well as differences with previous
work. Finally, we conclude with our findings and comment on the
results of the current manuscript in Sec.~\ref{sec:5}.

\section{Numerical scheme}
\label{sec:2}

Our approach to frustrated systems~\cite{anis_2,PT1,PT2} is based
on the parallel tempering (PT)
method~\cite{Swe86,Huku96,mariPL98,Newman99}. A PT protocol uses
an adequate number of lattice sweeps (usually Metropolis MC
steps~\cite{metro53}) followed by PT exchange moves between
neighboring temperatures. Simulations extend to a temperature
range, which depends on the linear lattice size $L$, and is
appropriate for the estimation of the critical properties of the
system. In other words, we generate MC data that cover several
finite-size anomalies of the system. The temperatures are selected
using a constant acceptance exchange
method~\cite{PT1,Sug99,earl05,bittner08,bittner11} and the
acceptance rate is chosen within the range $0.15 - 0.5$. This PT
practice is a quite general approach used, also by several other
authors~\cite{Sug99,earl05,bittner08,bittner11}, and has been
carefully implemented and compared to other alternative PT methods
in our recent papers~\cite{PT1,PT2}. To obtain the temperature
sequences for the production runs, we have used relatively short
preliminary runs to generate adequate MC data in the range of
interest. A simple histogram
method~\cite{Swendsen87,Newman99,LandBind00} was then used to
determine from the energy probability density functions the
temperatures, satisfying the constant acceptance exchange
condition~\cite{bittner11,PT1}. Since these temperature sequences
depend weakly on the disorder realization, we average over several
realizations to find an averaged sequence, which is subsequently
used in the extended runs.

For each case, corresponding to a certain set of $[p_{z} ; L]$
values, the PT scheme was carefully tested for equilibration.
Several independent runs of a large number of disorder
realizations were carried out in order to obtain, in the
temperature range of interest, the relevant disorder averaged
parameters $[Z]$, where $Z$ denotes a thermal average of some
thermodynamic quantity. We ensured that the temperature sequence
went well deep into the paramagnetic phase in order to avoid
entrapment and ensure equilibration. For moderately small lattice
sizes, the number of temperatures used were approximately $5-10$,
but for larger sizes we had to use more than $20$ temperatures,
even though we had adjusted the acceptance rate in a rather low
value, of the order of $0.15$. As expected, equilibration was much
easier in the paramagnetic and ferromagnetic phases, than in the
glassy phase. Furthermore, as sample-to-sample fluctuations are
substantially larger for glassy systems, not only the averaging
time, but also the number of random samples simulated had to be
accordingly adjusted.

In more detail, for the lower part of the phase diagram, $p_{z}
\in \{0.15, 0.25, 0.35, 0.40\}$, the signature of the F - P
transition was very clear. This can be observed from the
illustrations of Section \ref{sec:3B}, where the finite-size
anomalies of the system are illustrated. In these cases we have
used $6$ to $7$ different lattice sizes varying from $L=8$ to $32$
and $400$ to $1000$ disorder realizations. In particular, for the
case $p_{z}=0.25$ we simulated systems up to $32\times 32\times
32$ spins. This specific case was used as the representative of
the F - P line and a detailed FSS analysis was carried out and is
presented in Sec.~\ref{sec:3B}. In this particular case, the
corresponding ensembles of realizations varied from $1000$ samples
for $L=8$ to $400$ samples for the largest lattice $L=32$. Near
the multicritical point, a much larger set of realizations was
needed. For $p_{z}=0.425$, the number of realizations varied from
$7000$ for $L=8$ to $600$ for $L=24$. For the case $p_{z} =0.45$
we started with $L=8$ and $14000$ realizations going up to $L=20$
with $2400$ samples. Deep in the SG - P transition line, near the
symmetric point $p_{z}=0.5$, smaller system sizes were simulated.
For $L=6$ we used $24000$ samples for $p_{z}=0.475$ and $36000$
samples for $p_{z}=0.5$. We reached the size of $L=24$ with $720$
samples for $p_{z}=0.475$ and the size of $L=16$ with $6000$
samples for $p_{z}=0.5$. Further details on our simulations will
be be given in the sections.

\section{Finite-size scaling analysis and results}
\label{sec:3}

We start this Section by outlining the FSS analysis
(Sec.~\ref{sec:3A}) used for the estimation of critical
parameters, and we also present, in subsections, estimations for
the transition lines and the corresponding critical behavior of
the present anisotropic model. All the results shown below refer
to the disorder-averaged thermodynamic quantities, as explained in
more detail below. For instance in Sec.~\ref{sec:3B} a detailed
study is presented for the case $p_{z} = 0.25$. Similar analysis
has also been performed for several other values of $p_{z}$ along
the global phase diagram, but the details of some cases are
omitted for reasons of brevity. However, a full summary of results
for the critical temperatures and the correlation length's
exponent for all the values of $p_{z}$ simulated is given in
Table~\ref{tab:1} at Sec.~\ref{sec:3D}.

\subsection{Finite-size scaling framework}
\label{sec:3A}

Among other alternatives, we attempt here to estimate the critical
properties applying the traditional FSS route, in which one
observes the scaling laws of several finite-size anomalies of the
system. The corresponding peaks of specific heat $C$, magnetic
susceptibility $\chi$, and the peaks of logarithmic derivatives of
the order parameter with respect to the inverse temperature
$K=1/T$ will be used. For a disordered system, we assume that the
averaged over disorder $[Z]$ parameter of the system is the
relevant observable, where $Z$ denote the thermal average of some
thermodynamic quantity, such as the ones mentioned above.
Furthermore, the standard scaling power laws, commonly used for
the thermal averages of a pure system, are assumed to apply to the
corresponding $[Z]$ observable of the disordered system.
Finite-size anomalies are now defined with the help of $[Z]$.

In this scaling scheme, the maxima of the specific heat are
assumed to follow the scaling law $[C_{L}]^{\ast}\sim
L^{\alpha/\nu}$, whereas the peaks of the magnetic susceptibility
are expected to scale as
\begin{equation}
\label{eq:chi} [\chi_{L}]^{\ast}\sim L^{\gamma/\nu}.
\end{equation}
Additionally, the peaks of the logarithmic derivative of power
$n=2$ of the order parameter with respect to the inverse
temperature $K=1/T$, defined as~\cite{ferrenberg91},
\begin{equation}
\label{eq:moment} \frac{\partial \ln \langle
M^{2}\rangle}{\partial K}=\frac{\langle M^{2}H\rangle}{\langle
M^{2}\rangle}-\langle H\rangle
\end{equation}
are assumed to follow a power-law behavior with the correlation
length exponent of the form
\begin{equation}
\label{eq:moment_scaling} \left [ \left ( \frac{\partial \ln \langle
M^{2}\rangle}{\partial K} \right ) _{L} \right ]^{\ast} \sim L^{1/\nu}.
\end{equation}
\begin{figure}[htbp]
\includegraphics*[width=8.0 cm]{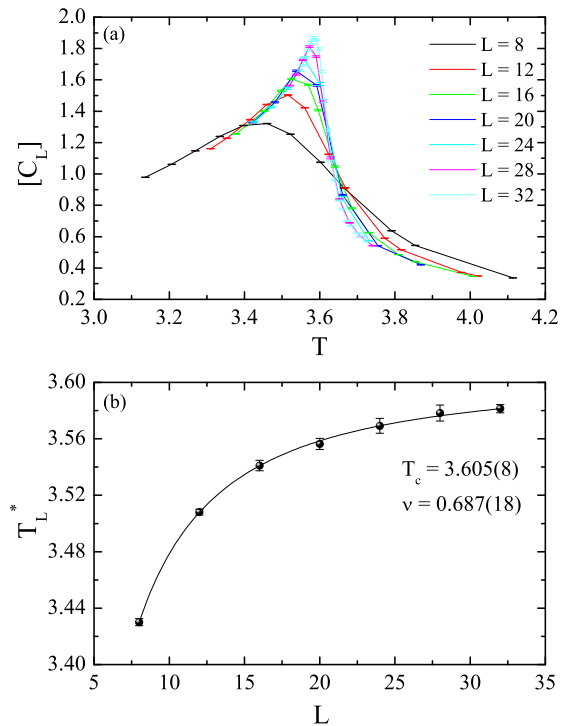}
\caption{\label{fig:spec_heat_scaling} (color online) (a)
Specific-heat curves as a function of the temperature. (b) Shift
of the corresponding pseudocritical temperatures obtained from
panel (a) where the specific heat attains its maximum.}
\end{figure}
The shift behavior of these finite-size anomalies $[Z]^{\ast}$ is
used for the estimation of critical temperatures and the
correlation length exponent by applying the standard fitting
formula for second-order phase transitions
\begin{equation}
\label{eq:shift} T_{L}^{\ast} = T_{\rm c} + b L^{-1/\nu}.
\end{equation}

For the present study we have used two different order parameters:
(i) the usual magnetization $M$, which is appropriate for
describing transitions that involve the ferromagnetic phase,
defined as
\begin{equation}
\label{eq:M} M=\frac{1}{N}\sum_{i=1}^{N}s_{i},
\end{equation}
where $s_{i}$ is the spin variable and $N$ the number of lattice
sites and (ii) the spin-glass overlap order parameter $q$,
appropriate for the SG - P transition (also for the F - P
transition), defined respectively as
\begin{equation}
\label{eq:q} q = \frac{1}{N}\sum_{i=1}^N s_i^{\alpha} s_i^{\beta},
\end{equation}
where $s_i$ denotes the spin of the site $i$ and $\{ \alpha,
\beta\}$ represent two replicas of the same disorder realization.
For the disordered system the corresponding, averaged over
disorder, order parameter is obtained from the thermal averages of
Eqs.~(\ref{eq:M}) and (\ref{eq:q}).

\begin{figure}[htbp]
\includegraphics*[width=8.0 cm]{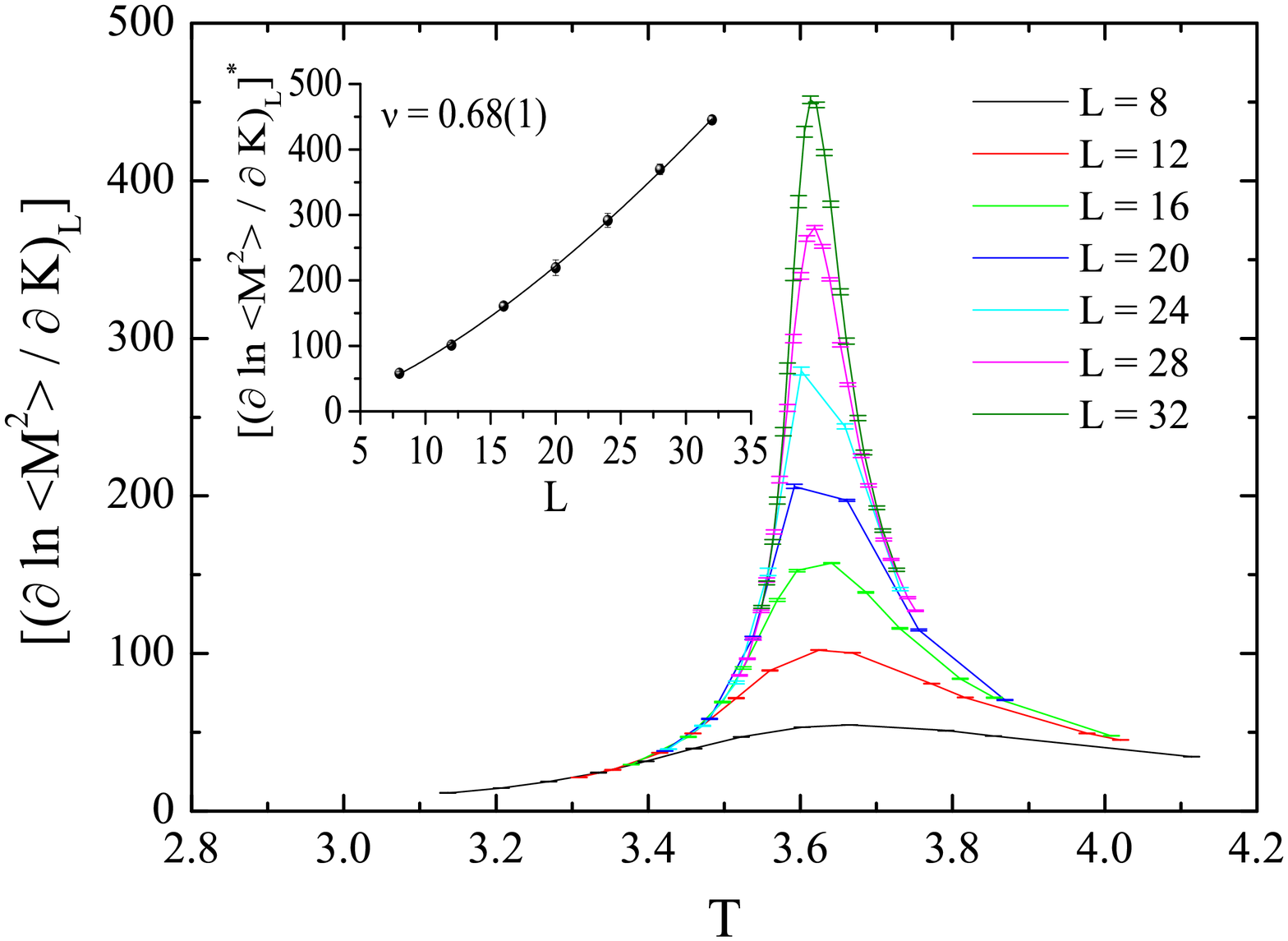}
\caption{\label{fig:ropes} (color online) Temperature dependence
of the second moment of the logarithmic derivative of the order
parameter with respect to the inverse temperature (main panel) and
the corresponding FSS behavior of the peaks (inset).}
\end{figure}

Following common practice
\cite{FG_1,Hase10,hartmann_sggs,balle_sg,JankeBook,katz_pg}, we
use a definition for the Binder cumulant~\cite{LandBind00} of $Z$
($M$ or $q$), based on the disorder averaged moments
\begin{equation}
\label{eq:cum}
 U_{Z} = 1-\frac{[\langle Z^{4}\rangle]}{3[\langle
 Z^2\rangle]^{2}}.
\end{equation}
The crossings of the order-parameter's fourth-order Binder
cumulant $U_{M }$ or $U_{q}$ have been used as an alternative
method in estimating the phase-diagram points. Furthermore,
another route that provide us with complementary estimates of both
the critical temperature and the exponent $\nu$ is that of data
collapse~\cite{LandBind00}. The application of this method is
carried out assuming a scaling hypothesis of the form
\begin{equation}
\label{eq:collapse} U_{Z} \approx
f[(T-T_c)L^{1/\nu},
\end{equation}
with $Z$ the above described order parameters. For the fitting
procedure we have used \emph{autoScale}, a program that performs a
FSS analysis for given sets of simulated data~\cite{collapse}. The
program implements the above scaling assumption which is expected
to hold best close to the critical point and optimizes an initial
set of scaling parameters that enforce a data collapse of the
different sets. The optimum data collapse, achieved by the
minimization procedure of the scaling parameters via the downhill
simplex algorithm, is carried out by finding a fair compromise
between a (rather small) value of a $\chi^{2}$/dof-like quantity
and a reasonably large interval on the re-scaled abscissa.

\subsection{Ferromagnetic - Paramagnetic Transition}
\label{sec:3B}

\begin{figure}[htbp]
\includegraphics*[width=8.0 cm]{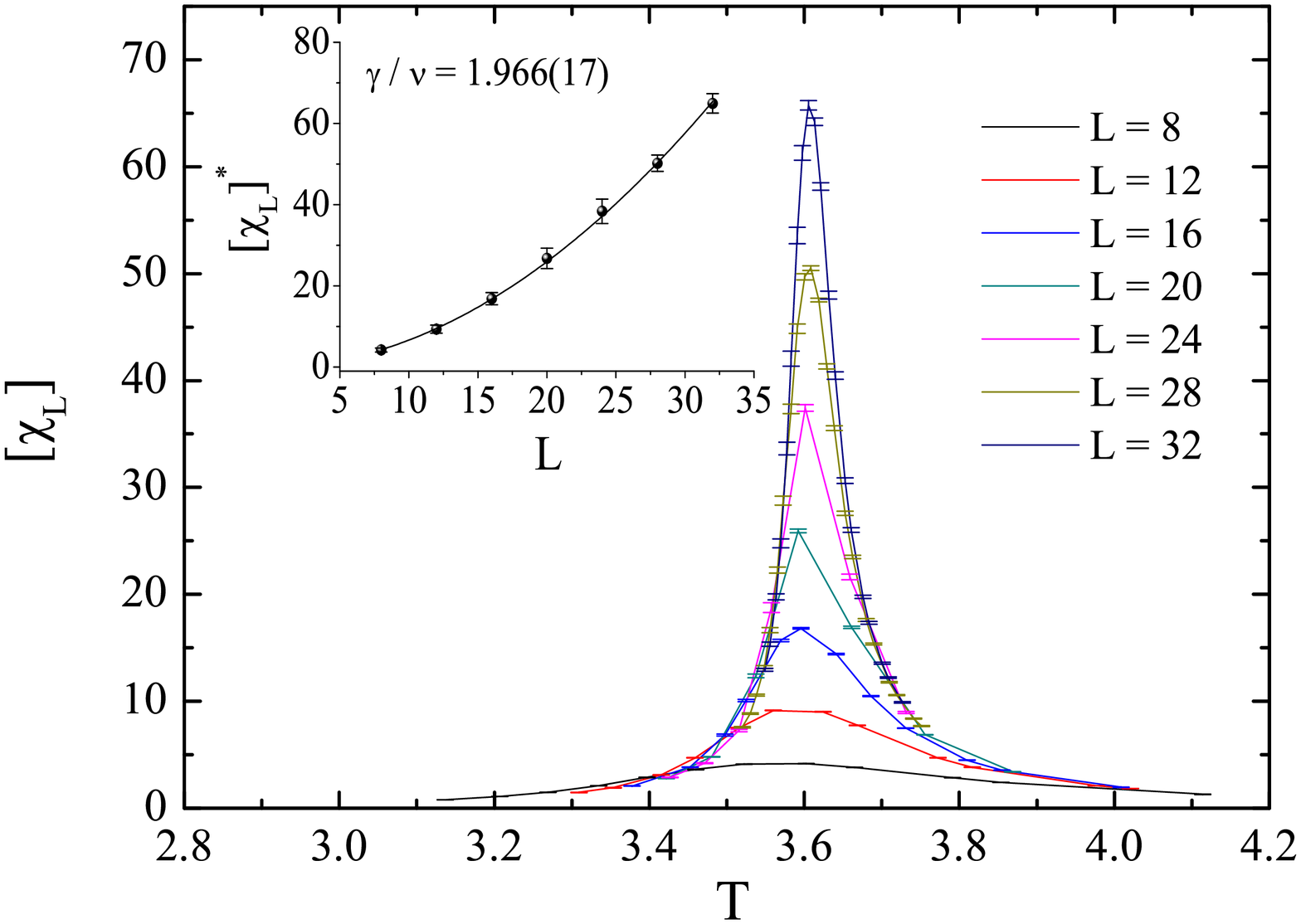}
\caption{\label{fig:magn_susceptibility} (color online)
Illustration of the magnetic susceptibility as a function of the
temperature (main panel) and the FSS of the peaks (inset).}
\end{figure}

Our main interest is the identification of the universality class
of the model and this is, in general, characterized mainly from
the critical exponent $\nu$ of the correlation length. Thus, we
shall focus our analysis on the estimation of this exponent.
Additional tests will include the estimation of the critical
temperature via various FSS schemes, reflecting also the accuracy
of our numerical data, as well as the estimation of the magnetic
exponent ratio $\gamma/\nu$.

We start our analysis with Fig.~\ref{fig:spec_heat_scaling},
illustrating the scaling behavior of the specific heat. In
particular, the upper panel (a) of this figure illustrates the
specific-heat curves $[C_{L}]$ as a function of the temperature
for the complete set of system sizes studied. The observed shift
behavior of the specific-heat peaks is quantified in the
corresponding lower panel (b), where we study the FSS of the
relevant pseudocritical temperatures $T_{L}^{\ast}$, i.e., the
temperatures where the specific heat attains its maximum. Applying
the standard fitting formula for second-order phase
transitions~(\ref{eq:shift}) on the numerical data we get the
estimates $T_{\rm c} = 3.605(8)$ and $\nu=0.687(18)$ for the
critical temperature and the correlation length's exponent,
respectively, the latter indicating that the model shares the
random Ising universality class, for which accurate estimates are
$\nu=0.6837(53)$~\cite{balles_rdi}, $0.683(3)$~\cite{hasen_fp},
and $0.6835(25)$~\cite{anis_2}.

A further verification of the value of the critical exponent $\nu$
is provided via the FSS analysis of the peaks of the logarithmic
derivative of the power $n=2$ of the order parameter with respect
to the inverse temperature $K=1/T$, as shown in
Fig.~\ref{fig:ropes}. In particular, in the main panel of this
figure we illustrate the temperature dependence of the second
moment [Eq.~(\ref{eq:moment})] for the whole spectrum of lattice
sizes studied and in the corresponding inset the FSS of the peaks.
These are expected to scale as Eq.~(\ref{eq:moment_scaling}) and
the power-law fitting, shown by the solid line, provides an
estimate $\nu=0.68(1)$ for the exponent, in agreement with the
values given above.

\begin{figure}[htbp]
\includegraphics*[width=8.0 cm]{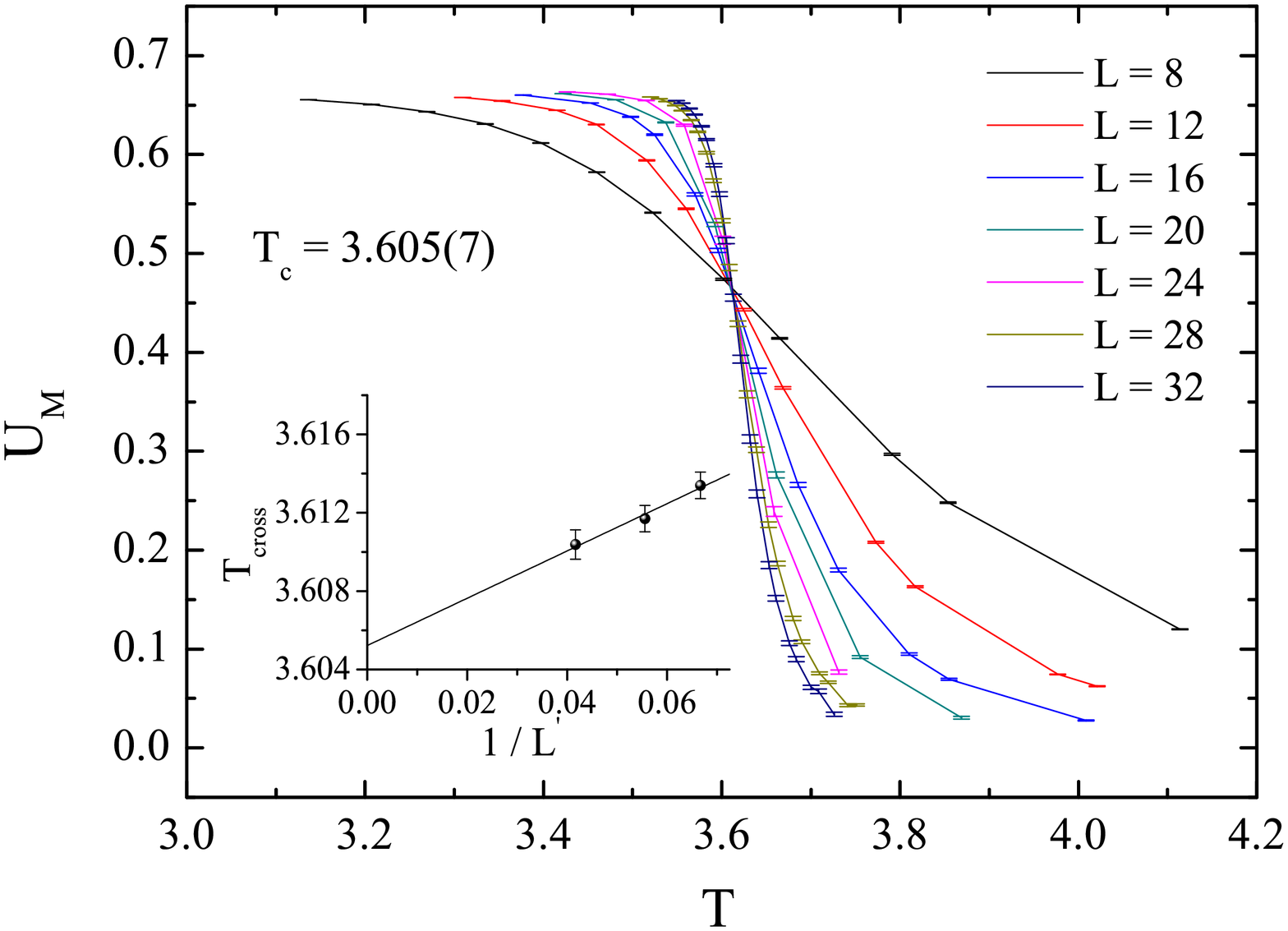}
\caption{\label{fig:crossings} (color online) Crossings of the
order-parameter's fourth-order Binder cumulant (main panel) and
infinite limit size extrapolation of the crossing points (inset).}
\end{figure}

We conclude our standard FSS analysis with the magnetic properties
of the model, and in particular with the estimation of the
magnetic exponent ratio $\gamma/\nu$ through the FSS of the
magnetic susceptibility. In the main panel of
Fig.~\ref{fig:magn_susceptibility} we present the temperature
dependence of the magnetic susceptibility and the FSS of the
corresponding peaks in the inset. The solid line shows a power-law
fitting of the form~(\ref{eq:chi}) using the complete
lattice-range spectrum. The outcome of the fitting provides an
estimate $1.966(17)$ for the exponent ratio $\gamma/\nu$, in good
agreement to the best-known literature estimates for the random
Ising model which slightly vary around the value
$1.965(6)$~\cite{balles_rdi,berche-04,fytas-10}.

The above estimates are now further verified from a different
approach, see Fig.~\ref{fig:crossings}, where the crossings of the
order-parameter's fourth-order Binder cumulant $U_{M}$ are
illustrated. Although it is clear that the curves cross around the
value $3.6$ which marks the location of the critical temperature
of the system, a more refined analysis (see the inset) using the
infinite-limit size extrapolation $L'\rightarrow \infty$, where
$L'=1/(L_{1}+L_{2})$, of the crossing points $T_{\rm cross}$ of
the pairs $(L_{1},\; L_{2})=(L,\; 2L)$ gives an estimate of
$T_{\rm c} = 3.605(7)$ for the critical temperature.

\begin{figure}[htbp]
\includegraphics*[width=8.0 cm]{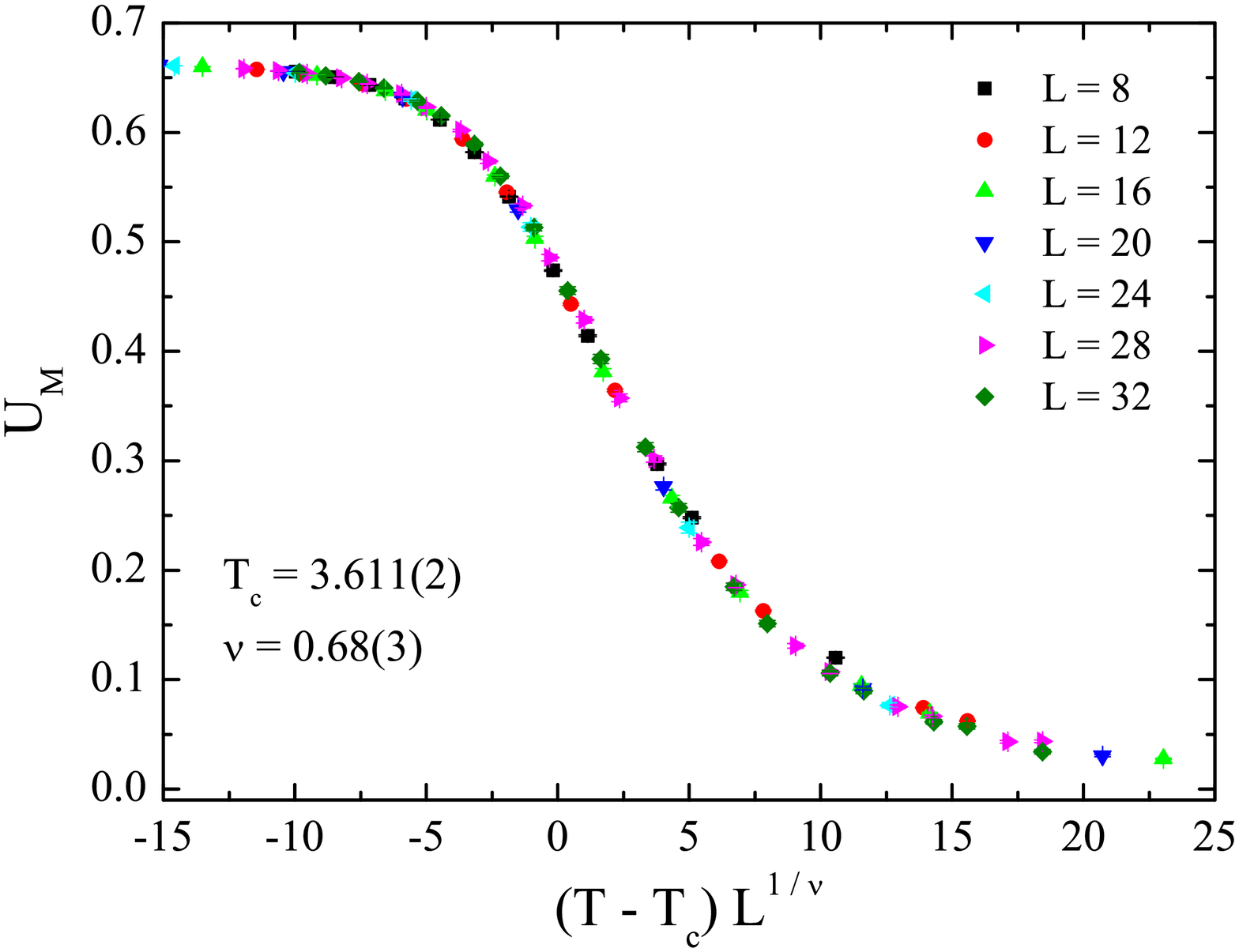}
\caption{\label{fig:collapse_V_M} (color online) Collapse of the
order-parameter's fourth-order Binder cumulant of $M$.}
\end{figure}

Last but not least, we provide complementary estimates of both the
critical temperature and the exponent $\nu$, via the method of
data collapse~\cite{collapse}, as already discussed above in
Sec.\ref{sec:3A}. In the current case, the optimum data collapse
for the order-parameter's fourth-order Binder cumulant $U_{M}$ is
shown in Fig.~\ref{fig:collapse_V_M} and the resulting value for
the critical temperature $T_{\rm c} = 3.611(2)$ is in agreement
with the previous estimates of Fig.~\ref{fig:crossings} from the
crossings of the same thermodynamic quantity. Additionally, the
estimate $\nu=0.68(3)$ for the critical exponent of the
correlation length is also in agreement with the previously
obtained estimates.

Similar collapse attempt was performed for the Binder cumulant of
the overlap order parameter $U_{q}$ and both critical estimates
are presented in Table~\ref{tab:1}. Further cases for the lower
part of the phase diagram, $p_{z}\in \{0.15, 0.35, 0.40\}$, were
considered and the estimates are also in Table~\ref{tab:1}. For
the values $p_z=0.425$ and $p_z=0.45$ we observe a considerable
deviation in the estimated critical temperatures obtained from the
corresponding collapse of $U_{M}$ and $U_{q}$, indicating that we
are close to the multicritical point, in which the magnetization
ceases to operate as the appropriate order parameter of the
system. In particular, at $p_{z}=0.45$ we observe a significant
deviation also in the estimate of the correlation length exponent.
In this case the exponent produced from the collapse of the
magnetization data is $1/ \nu=0.86(12)$, whereas that produced
from the overlap data is $1/ \nu=0.67(12)$. The last value is
close to $1/ \nu=0.61(2)$ given by Hasenbusch \emph{et
al.}~\cite{hasen_mcp}, confirming that $p_{z}=0.45$ is close to
the multicritical point.

\subsection{Spin Glass - Paramagnetic Transition} \label{sec:3C}

From the closing remarks of the previous section it is expected
that for the present model the multicritical point is close to
$p_{z}=0.45$. The data for the F - SG transition line in
Sec.~\ref{sec:3D} are also in agreement with this estimate. Thus,
two values $p_{z}=0.475$ and $p_{z}=0.5$ of our study belong to
the SG - P transition line. For brevity we omit the details of our
analysis for the first value, and we concentrate here only in the
most interesting symmetric, $p_{z}=0.5$, case including a relevant
ground-state study.

\begin{figure}[htbp]
\includegraphics*[width=8.0 cm]{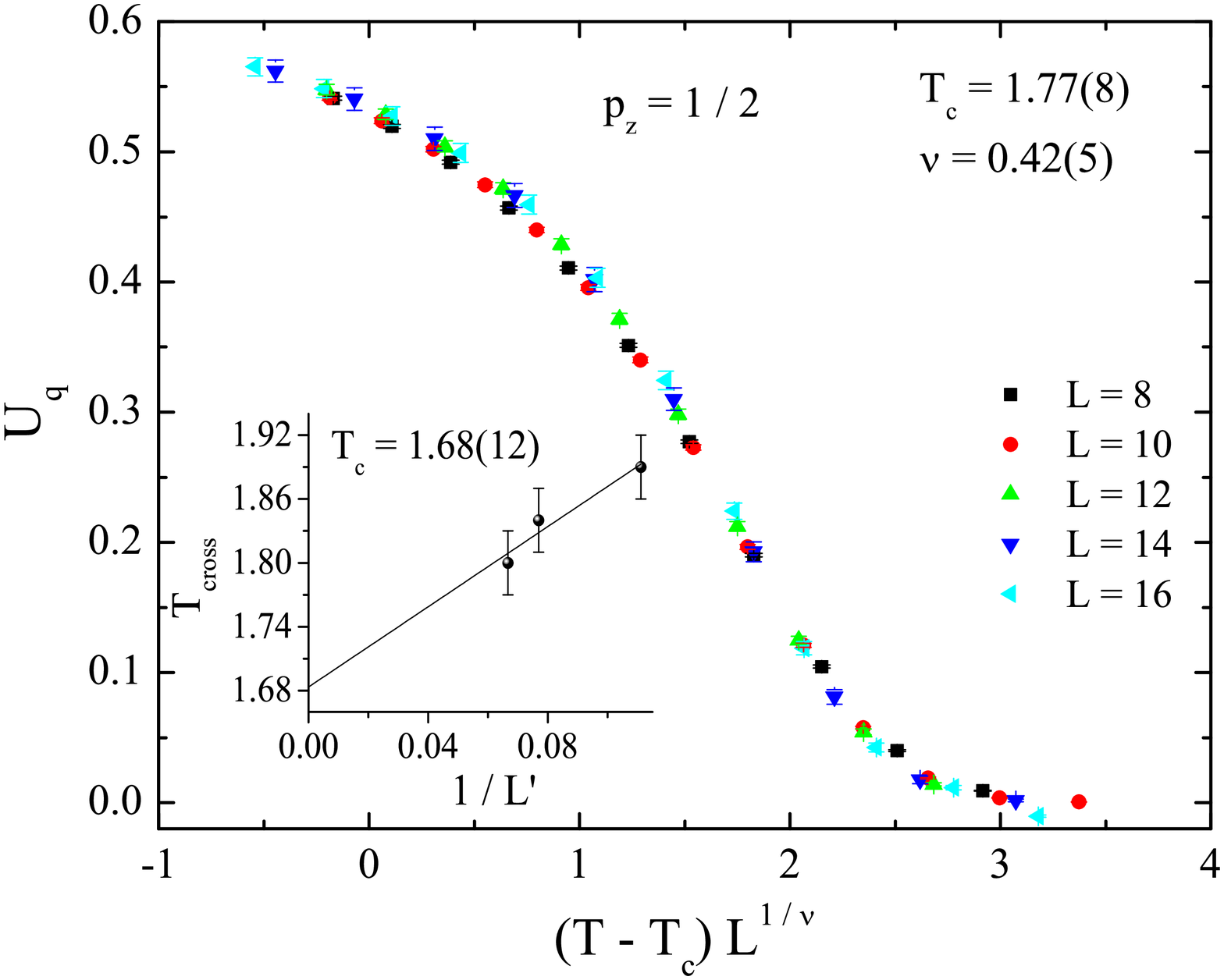}
\caption{\label{fig:collapse05} Critical temperature at the
symmetry axis $p_{z}=1/2$. The collapse of the overlaps's
fourth-order Binder cumulant with the method outlined in
Sec.~\ref{sec:3} gives $T_{\rm c}=1.77(8)$, while the
infinite-limit size extrapolation of $U_q$ crossings, illustrated
in the inset, gives $T_{\rm c}=1.68(12)$. Details are provided in
Sec.~\ref{sec:4}}.
\end{figure}

In the vicinity of the symmetric case $p_{z}=1/2$ at low enough
temperatures, the presence of the spin-glass phase was clearly
detected. The first indication was that the magnetization in this
regime was not an appropriate order parameter (see also below the
discussion for the $T=0$ magnetization order). Furthermore, at
$p_z=0.475$ and $p_z=1/2$ (see Fig.~\ref{fig:collapse05}), the
data collapse analysis of $U_{q}$ produced critical exponent
values $\nu=2.370(14)$ and $2.381(8)$ respectively, which are
compatible to the estimated values of the SG - P transition found
in the literature, namely the values $2.72(8)$~\cite{campb_sg},
$2.15(15)$~\cite{balle_sg}, $2.22(15)$~\cite{jorg}, and the recent
estimates $2.45(15)$~\cite{hasen_sg} and $2.39(5)$~\cite{katz_pg}.
Similar results were also found in our recent study of the
anisotropic case $\{p_{xy}\leq 0.4 ; p_z=0\}$, where $\nu$ was
estimated $2.424(14)$~\cite{anis_2}.

The collapse of the Binder's cumulant of the overlap order
parameter, illustrated in the main panel of
Fig.~\ref{fig:collapse05}, gives an estimate of the order of
$T_{\rm c}=1.77(8)$ for the critical temperature. We have also
performed a linear extrapolation of the crossings points of
$U_{q}$, as illustrated in the inset of this figure. The fit is
performed in $1/L'$, with
$L'=\{\frac{(L_8+L_{10})}{2},\frac{(L_{12}+L_{14})}{2},\frac{(L_{14}+L_{16})}{2}\}$.
The resulting estimate is now $T_{\rm c}=1.68(12)$ in accordance
with the value produced by the data collapse method. Therefore,
the critical temperature of the longitudinal anisotropic model
(denoted by $T_{B_{xy}}$) for the symmetric case is estimated to
be considerably higher from the corresponding critical temperature
of the isotropic model. Let us recall that the critical
temperature of the isotropic model, denoted here by $T_{B}$, is of
the order of $T_{B}=1.109(10)$~\cite{hasen_sg}, and the
corresponding temperature for the transverse anisotropic model
$\{p_{xy} \leq 1/2;p_z=0\}$, denoted here by $T_{B_{z}}$, was
estimated in the same range $T_{B_{z}}=1.111(25)$~\cite{anis_2}.
The striking coincidence of points \textmd{B} and
\textmd{B}$_{z}$, was further supported by the detailed
ground-state study of Ref.~\cite{PT1}.

\begin{figure}[htbp]
\includegraphics*[width=8.0 cm]{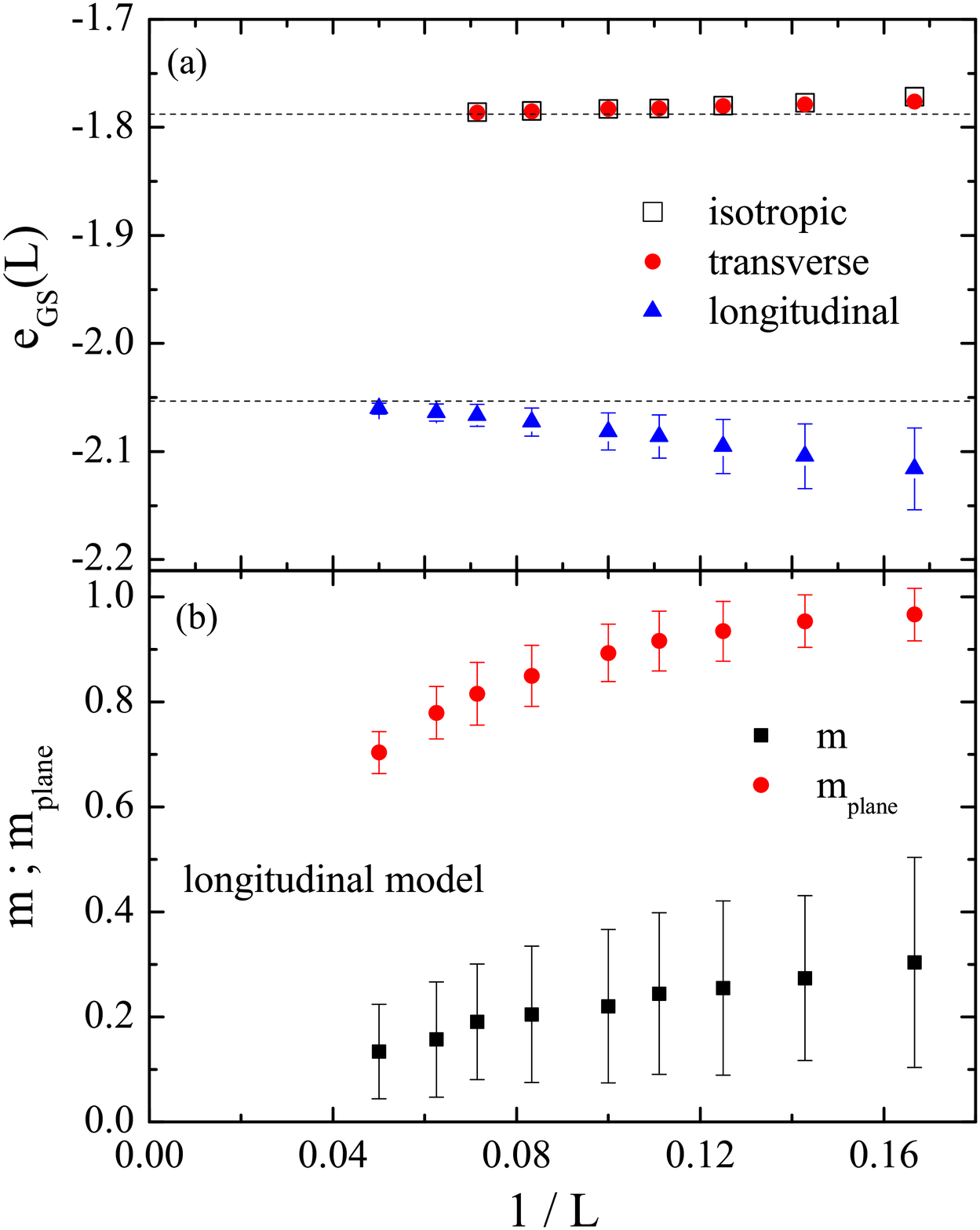}
\caption{\label{fig:GS} (a) Behavior of ground-state energies per
site $e_{\rm GS}$ for the $d=3$ EA model and the two anisotropic
versions. (b) Size-decline of the $T=0$ magnetization order in the
bulk and $xy$-planes of the present longitudinal anisotropic
model. In both panels error bars denote sample-to-sample
fluctuations.}
\end{figure}

In relevance to the above discussion, we now present in
Fig.~\ref{fig:GS} the finite-size behavior of the ground-state
energy per site($e_{\rm GS}=E_{\rm GS} / N$) for both anisotropic
versions of our studies together with the known behavior of the
isotropic $d=3$ EA model. Evidently, the clear difference in the
limiting value of $e_{\rm GS}$ reflects the considerably higher
critical temperature of the longitudinal anisotropic model. The
known asymptotic estimations for the isotropic model are $e_{\rm
GS}(L\rightarrow \infty) = -1.7863(4)$~\cite{pal96}, and
$-1.7876(3)$~\cite{hart97}, and the second value is indicated by
the dashed line in Fig.~\ref{fig:GS}. For the present longitudinal
anisotropic model the dashed line indicates the estimate $e_{\rm
GS}(L\rightarrow \infty)=-2.0533(8)$. In the second part of this
figure, we also show the size-decline of the $T=0$ magnetization
order in the bulk $m = [\langle |M|\rangle]$. In the same figure
an alternative order parameter is used based on the magnetization
of the $xy$ planes, defined as $m_{\rm plane} = [\langle \sum_{\rm
planes} |M_{\rm plane}|\rangle]$. Apparently, this pseudo-ordering
(more pronounced for $m_{\rm plane}$) is only a finite-size effect
that keeps also the ground-state energy in much lower values. The
data shown in Fig.~\ref{fig:GS} were obtained via the PT algorithm
using a practise analogous to that detailed in Ref.~\cite{PT1}.
The ensemble of disorder realizations varied from $15\times 10^3$
for $L=6$ to $100$ for $L=20$.

We will now try to give an intuitive explanation of the above
observations concerning the critical temperatures at the symmetric
cases of the above spin-glass models. We suggest that, at these
symmetry points - $p=p_{xy}=p_{z}=1/2$ - the critical temperatures
$T_{B}$, $T_{B_{z}}$, and $T_{B_{xy}}$, are determined by some
simple global frustration features of the models. First, let us
assume that the coincidence of $T_{B}$ and $T_{B_z}$ observed in
our earlier papers~\cite{anis_2,PT1} is due to the fact that the
density of the elementary squares of the lattice, denoted by $f$,
which cannot simultaneously satisfy all their bonds (frustrated
elementary squares or plaquette) is equal to $f=1/2$ for both
models. This is also a plausible explanation of the apparent match
in their ground-state energy found in Ref.~\cite{PT1}(see also
Fig.~\ref{fig:GS}). For the current model, at the point $p_z=1/2$,
the corresponding density is $f=1/3$, giving a lesser amount of
frustration, thus producing a much higher $T_{\rm c}$, according
to the above hypothesis.

Further global frustration features of the models can be invoked
in order to support the above argument. As an example, let us
consider the class of completely unfrustrated bonds. These type of
bonds ($B_0$) on the simple cubic lattice have zero frustrated
plaquette attached to them and one can define the corresponding
densities $r({B_0})$ as fractions of the total number of bonds.
Then, it is again easily seen that for both the isotropic model at
$p=1/2$ and also for the transverse anisotropic model
$\{p_{xy}=1/2 ; p_z=0\}$, $r({B_0})=(1/2)^4=0.0625$, whereas for
the longitudinal anisotropic model this fraction is
$r({B_0})=[2(1/2)^4+(1/2)^2]/3=1/6=0.166\cdots$. Thus, the
longitudinal anisotropic model carries weaker frustration
features, leading to higher critical temperature. Similar
arguments can be applied for all five classes of bonds, i.e,
$B_{0}$, $B_{1}$, $B_{2}$, $B_{3}$, and $B_{4}$ on the simple
cubic lattice or their combinations, strengthening the hypothesis
that $T_{\rm c}$ is mainly determined by simple global frustration
features of the models.

\subsection{Ferromagnetic - Spin Glass Transition} \label{sec:3D}

The F - SG is the least investigated of the transition lines. A
recent comprehensive FSS analysis for the isotropic EA model was
performed by Ceccarelli \emph{et al}.~\cite{FG_1}. These authors
found a new universality class by determining two points of the F
- SG line corresponding to the temperatures $T=0.5$ and $T=1$.
These two phase-diagram points, when compared to previous
literature estimates for the multicritical and the $T=0$ F - SG
critical point, clearly supported the earlier proposed reentrant
behavior of the line. The extensive simulations of
Ref.~\cite{FG_1} involved lattice sizes within the range $L=4 -
20$, using many thousands of disorder realizations, even for
$L=20$.

\begin{figure}[htbp]
\includegraphics*[width=8.0 cm]{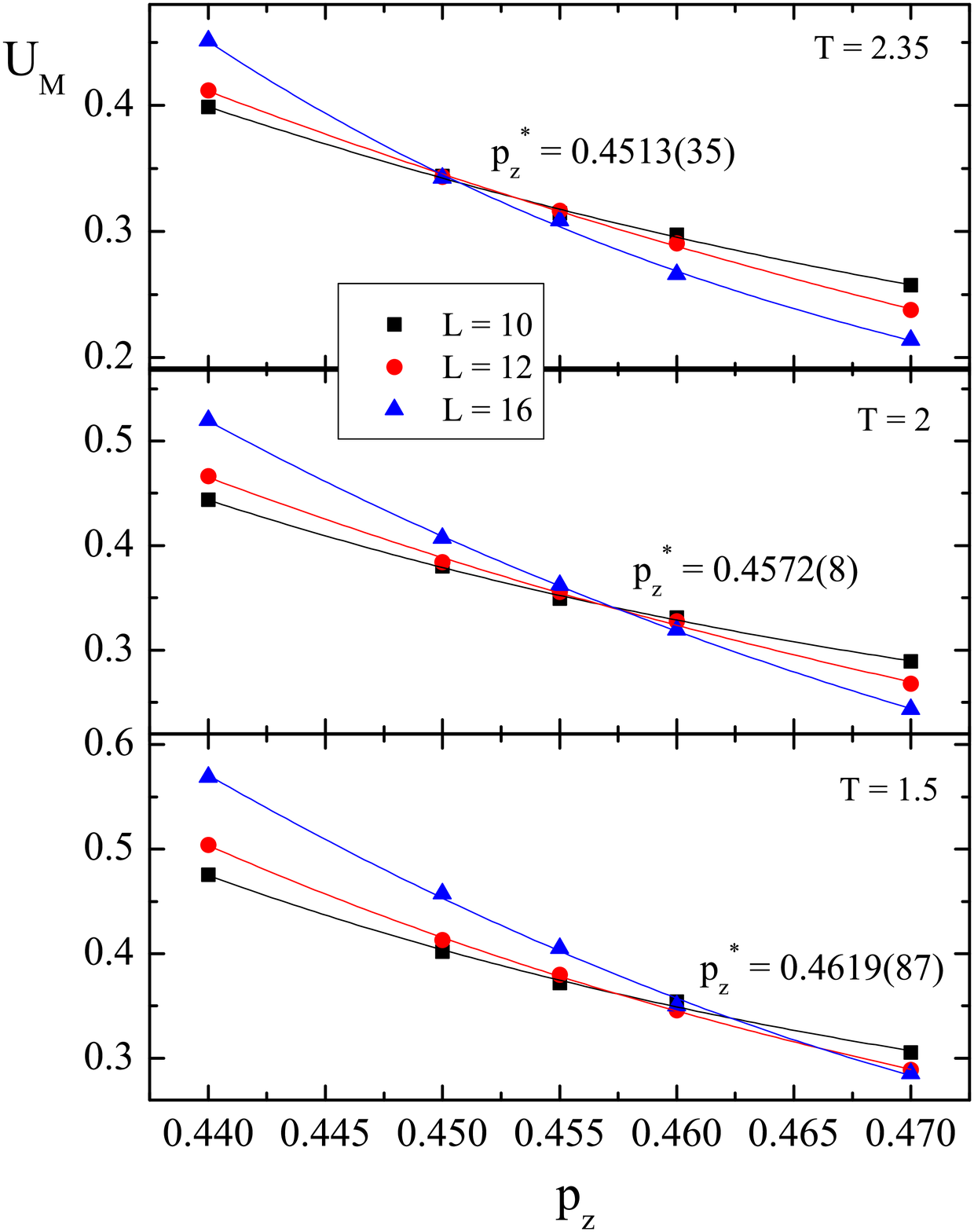}
\caption{\label{fig:multi} Illustration of the crossing behavior
of the magnetization's fourth-order Binder's cumulant versus
$p_{z}$ for the linear sizes $L=10$, $12$, and $L=16$ for three
relevant temperatures outlined. The resulting F - SG transition
line appears to be forward, according to the finite-size
approximation shown.}
\end{figure}

Our attempt here to estimate the F - SG line for the present
anisotropic EA model is not as extensive, we considered lattice
sizes $L=10$, $12$, and $L=16$, with $8000$, $3000$, and $2000$
samples, respectively. The simulations were carried out for
several values of $p_{z}$, close to the multicritical value
$0.45$. Figure~\ref{fig:multi} illustrates the behavior of the
fourth-order Binder's cumulant of the magnetization $U_{M}$. The
temperature sequence for each $p_{z}$ case went well deep into the
paramagnetic phase ($T>2.6$) in order to avoid entrapment and
ensure equilibration. While in most cases the minimum temperature
was $T=1.5$, a very low temperature, namely $T=0.45$, was used in
some production runs in order to record ground-state properties.

At $T=2.6$ in the range $p_{z}=0.44-0.47$ for linear sizes $L=10$
and $12$, $U_{M}$ does not show crossing and goes to zero as the
size increases, indicating that we are in the paramagnetic region.
Below the phase-diagram boundary we find crossing points as
illustrated in Fig.~\ref{fig:multi} for three temperatures, namely
$T=1.5$, $2$, and $T=2.35$, corresponding to crossing
probabilities $p_{z}=0.4619(87)$, $0.4572(8)$, and
$p_{z}=0.4513(35)$, respectively. The above were determined by the
mean value of the crossings of the lines produced by second order
polynomial fittings on the simulated points. Note that these
values are rough estimates of the asymptotic results, since no
extrapolation has been attempted. The F - SG transition line
appears, at least according to the present finite-size
approximation, to be forward and not reentrant, as shown for the
isotropic model by Ceccarelli \emph{et al}.~\cite{FG_1}.

\begin{figure}[htbp]
\includegraphics*[width=8.0 cm]{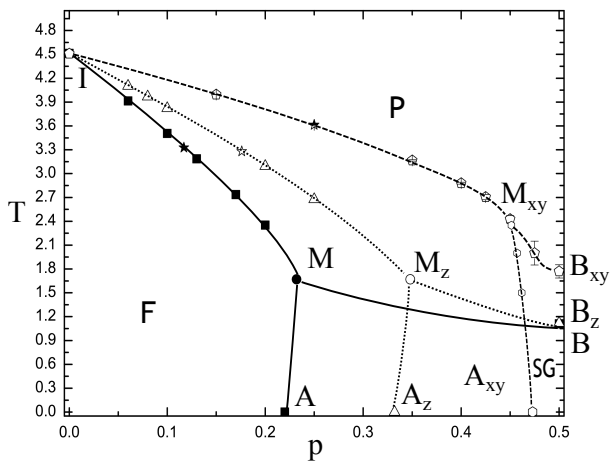}
\caption{\label{fig:phd}Global phase diagrams. Transition lines
separate ferromagnetic (F), spin glass (SG), and paramagnetic (P)
phases, and the multicritical points are denoted by \textmd{M}.
Solid lines and full symbols illustrate the phase diagram of the
isotropic EA model. F - P line points are taken from
Ref.~\cite{hasen_fp}, the multicritical point \textmd{M} from
Ref.~\cite{hasen_mcp}, the zero-temperature F - SG point
\textmd{A} from Ref.~\cite{hartmann_sggs}, and the symmetric SG -
P point \textmd{B} from Ref.~\cite{hasen_sg}. The dotted lines and
open symbols correspond to the transverse anisotropic model
$\{p_{z} = 0 ; p_{xy}\leq 1/2\}$ described in detail in
Ref.~\cite{anis_2}. Finally the dashed lines and open symbols
correspond to the present longitudinal anisotropic model
$\{p_{z}\leq 1/2 ; p_{xy}=0\}$.}
\end{figure}

From the above discussion, the F - SG transition line for the
present longitudinal anisotropic model appears to be forward,
although an uncertainty remains because of the rather limited
finite-size data at hand. However, the shifts of the crossings
introduced by adding $L=16$ reinforces our claim. This behavior,
if verified, could indicate to a link between frustration and the
slope of the transition lines, since the present model has quite
different frustration features close to the multicritical point.
Thus, a need for a more extensive FSS analysis in this regime
remains.

By using two estimates for the phase diagram, close to the
multicritical point for the F - P line and the above three
estimates of the F - SG transition line we find, using linear
fittings, $p_{z}=0.4509(5)$ and $T=2.42(3)$ for the multicritical
point. The $T=0$ F - SG critical point found by the linear
fittings based on the above three estimates is $p_{z}=0.48(1)$,
which is not far from the value obtained by a collapse of $U_M$
performed on our restricted ground-state data ($p_{z}=0.47(14)$).
These estimates indicate a clear forward behavior.

Concluding the above discussion, it is also useful to recall that
a forward ferromagnetic spin-glass line has been illustrated for a
transverse anisotropic model with ratio of interactions
$J^{z}/J^{xy}=0.5$ (see upper right panel of Fig.3 in
Ref.~\cite{anis_1}). For the same ratio of interactions, the
authors of Re.~\cite{anis_1} also reported a very narrow
spin-glass phase for the longitudinal anisotropic version of the
model without commenting on the forward or reentrant character of
this line (see the lower left panel of Fig.3 in
Ref.~\cite{anis_1}). The origin of these findings, including those
of the present paper, is not diaphanous and this makes the problem
even more interesting.

\section{Phase diagram and summary of estimates}
\label{sec:4}

In this Section we present the global phase diagrams of the
following spin models: the well-known isotropic EA spin-glass
model~\cite{ea-model,BindY86,hasen_fp}, the transverse anisotropic
case $\{p_{z} = 0 ; p_{xy}\leq 1/2\}$ presented in our previous
work~\cite{anis_2}, and finally the present longitudinal
anisotropic model $\{p_{z} \leq 1/2 ; p_{xy}=0\}$. The three
models are symmetric around the $p$-, $p_{xy}$-, and
$p_{z}=1/2$-axis respectively, thus the $p, p_{xy}, p_{z} > 1/2$
parts of the diagrams are omitted. The phase diagrams are
illustrated in Fig.~\ref{fig:phd}. Point \textmd{I}, is the well
studied pure Ising model with $T_{\rm
c}=4.5115232(16)$~\cite{Hase10,camp_is,guida_is,Butera-02,deng_is,blote_is}.
The full star at $p=0.117(3)$ in the F - P line denotes the
improved model proposed in Ref.~\cite{hasen_fp}, where scaling
corrections are minimum. Open and half full stars represent
respectively points of the transverse anisotropic model
$p_{xy}=0.176$~\cite{anis_2} and $p_{z}=0.25$ of the present
longitudinal anisotropic model. Note that, the $x$-axis labelled
as $p$, represents the probability $p$ for the isotropic model,
$p_{xy}$ for the transverse anisotropic model, and $p_{z}$ for the
present longitudinal anisotropic model. Further details are given
in the caption of Fig.~\ref{fig:phd}.

The phase diagrams of the isotropic and transverse anisotropic
models were discussed in Refs.~\cite{anis_2,PT1}, where the most
striking feature of the coincidence of points \textmd{B} and
\textmd{B}$_{z}$, was supported by the detailed ground-state study
of Ref.~\cite{PT1}. As seen now from Fig.~\ref{fig:phd}, the
present model yields a SG - P point \textmd{B}$_{xy}$ at a
significantly higher temperature and the origin of this phenomenon
we believe stems from the different frustration features, already
discussed in Sec.~\ref{sec:3C}.

\begin{table}
\caption{\label{tab:1}Summary of estimates for the critical
temperature and the critical exponent $\nu$ obtained for all the
values of $p_{z}$ studied. For $p_{z}\leq 0.45$ two sets of
results are given, obtained from the data collapse method as
discussed in the main text, corresponding to the order-parameter's
$U_{M}$ (second and third column) and overlap's $U_{q}$ (fourth
and fifth) fourth-order Binder cumulant. For $p_{z} > 0.45$,
estimates only via the collapse of $U_{q}$ are given.}
\begin{ruledtabular}
\begin{tabular}{ccccc}
 & \;\;\;\;\;\;\;\;\;\;\;\;\;\;\;\;\;\;\;\; $U_{M}$ & &
\;\;\;\;\;\;\;\;\;\;\;\;\;\;\;\;\;\;\;\; $U_{q}$ &\\
\hline
$p_{z}$ & $T_{\rm c}$ & $1/\nu$& $T_{\rm c}$ & $1/\nu$ \\
\hline
0.15   & 3.993(2)    & 1.465(9)   & 3.995(2)   & 1.46(2)     \\
0.25   & 3.611(2)    & 1.464(12)   & 3.6128(9)   & 1.46(4)     \\
0.35   & 3.172(2)    & 1.43(9)   & 3.166(9)   & 1.46(4)     \\
0.4    & 2.90(2)   & 1.46(21)   & 2.880(12)   & 1.47(12)     \\
0.425  & 2.74(3)   & 1.46(42)   & 2.705(33)  & 1.46(5)     \\
0.45   & 2.41(12)   & 0.86(12)  & 2.43(5)  & 0.67(12)     \\
0.475  &             &           & 2.00(15)  & 0.42(8)     \\
0.5    &             &           & 1.77(8)  & 0.42(5)
\end{tabular}
\end{ruledtabular}
\end{table}

\section{Discussion and concluding remarks}
\label{sec:5}

In the present manuscript, we presented the phase diagram and
critical behavior of a further anisotropic case of the $\pm J$
three-dimensional Ising model on the simple cubic lattice. The
current spatially anisotropic $\pm J$ bond randomness was applied
only in the $z$ direction, whereas in the other two directions,
$xy$ - planes, a ferromagnetic exchange was implemented. The phase
diagram of the model under study was compared to that of the
corresponding isotropic one, as well as to the phase diagram of
the anisotropic case in which the $\pm J$ bond randomness was
applied only in the $xy$ - planes. The observed differences in the
global phase diagrams were critically discussed, assuming that
some global frustration features of the models determine the
critical temperature, especially at the symmetry points of the
models. As we have shown, the differences in frustration features,
at the symmetry point, give rise in the present model to a
considerable increase of the transition temperature. Furthermore,
our data for the ferromagnetic - spin glass transition line are
supporting a forward behavior in contrast to the reentrant
behavior of the isotropic model, and this is also proposed as an
effect of the frustration features of the model.

Our scaling analysis verified, once more, the interesting feature
of the irrelevance for the critical behavior of the spatial
anisotropy. This aspect of universality was also found in the
renormalization-group study of Ref.~\cite{anis_1}. The signs of
this universality can be directly identified by simply comparing
our critical estimates shown in Table~\ref{tab:1} to those of the
relevant literature. It is natural to expect that the general
universality, observed among these models, is a reflection of the
spatially stochastic character of the quenched disorder and the
frustration, although the differences in some global
characteristics may produce very interesting changes in the
corresponding phase diagrams. It will be useful, as a feature
task, to further understand the effects of the global frustration
features on both phase diagrams and also on the critical behavior
of these simple spin-glass models. Research in this direction is
currently under way.

\begin{acknowledgments}
A.M. acknowledges financial support from Coventry University
during a research visit at the Applied Mathematics Research
Centre, where part of this work has been completed.
\end{acknowledgments}

\end{document}